\documentclass[preprint]{aastex}

\def\ltsim{ \,{}^<_\sim\, }
\def\gtsim{ \,{}^>_\sim\, }
\shorttitle{Mass and Metallicity}
\shortauthors{W.~E.~Harris}

\begin{document}
\title{Cluster Mass and Metallicity Distributions: \\
  Reconstructing the Events During Halo Formation}
\author{William E.~Harris}
\affil{Department of Physics and Astronomy, McMaster University \\
Hamilton L8S 4M1 Canada}

\begin{abstract}
Globular clusters in most large galaxies are a mixture of
metal-poor and metal-rich (bimodal),
but the halo {\sl stars} are almost entirely {\sl metal-rich}.  
This and other lines of evidence argue that the metal-poor globular
clusters formed within widely distributed $10^8-10^9 M_{\odot}$
gas clouds (supergiant GMCs) during an early burst in  
which most of the gas was
ejected or unused till later rounds of star formation.

New simulations of the growth of pre-galactic potential wells in
the early universe now indicate that the initial power-law
form of the globular cluster mass distribution ($dN/dM \sim M^{-1.8}$)
is a miniature replica of the mass distribution of the SGMCs themselves,
which grow hierarchically in the CDM potential wells of large
protogalaxies.
\end{abstract}

\section{Introduction}

The simple goal of employing globular clusters 
as archaeological clues to galaxy formation 
has become increasingly complicated over the
past decade by the growing evidence thatf
globular cluster systems (GCSs) might have been strongly influenced by
major mergers and ongoing accretion events. 
Decoding the meaning of the
cluster metallicity distribution function (MDF) or the luminosity
distribution function (LDF or GCLF) has become more ambiguous.

The issue is made still more difficult by the realization that
the GCS is a {\sl highly biased and tiny subsample} of the old-halo
population:  the host galaxy is, {\sl by definition}, made up of
the field-halo stars, of which perhaps 0.3\% find themselves in
globular clusters at present (McLaughlin 1999).  
How typical is the MDF of the clusters
(e.g.) of the halo as a whole?  By studying the GCS in detail, are
we actually learning more about globular cluster formation processes
than about the larger problem of galaxy formation?

\section{NGC 5128:  A Case Study in MDFs}

Except for the Milky Way and a few galaxies in the Local Group,
it has not been possible to compare MDFs of the globular clusters
and halo stars {\sl within the same galaxy} in detail till recently;
and till now, the important gE type of galaxy has been out of reach.
We are just beginning to take some important steps in this direction.

NGC 5128, easily the closest giant E galaxy at $d \sim 4$ Mpc, is
beginning to emerge as a keystone object. 
A series of photometric studies 
(Soria et al. 1996; G.~Harris et al. 1999; Harris \& Harris 2000, 2001; 
Marleau et al. 2000) now give us a first
deep look into the true composition of a giant elliptical
on a {\sl direct}, star-by-star basis.  
Harris et al. (1999, 2000, 2001) give
the results of deep $HST$/WFPC2 $(V,I)$ photometry in three fields at 
projected radial distances from 8 kpc
out to 31 kpc from galaxy center.  The color-magnitude diagrams
of these ``pencil-beam'' samples of the halo reveal that
(a) more than 99\% of the stars can be interpreted as classically ``old'' 
red giants, with little or no contribution from younger AGB-type stars;
and (b) their color distribution is enormously broader than can
be explained by either photometric measurement scatter or internal
age spread, strongly indicating that 
a large star-to-star metallicity spread is present.

A dense grid of red-giant evolutionary tracks
(VandenBerg et al.~2000; Bertelli et al.~1994; see Harris \& Harris
2000 for detailed discussion and methodology) can be superimposed
on the color-magnitude diagram of the halo stars to transform
color $(V-I)_0$ into heavy-element abundance $Z$ and derive a 
first-order MDF (see Figure 1).

Most of the weight in the resulting MDF comes from
the brighter stars
in the diagram ($M_{bol} \ltsim -2.5$), where photometric
scatter is minimal and the model track separation is largest.
The overwhelming majority of the stellar population even in
the outer halo falls in the range $-1 <$ log $(Z/Z_{\odot}) < 0$.
There are {\sl almost no metal-poor stars},
nor are there many with above-Solar metallicities.
Similarly, {\sl small} fractions of low-metallicity stars
have been deduced for other E galaxies from population 
synthesis of their integrated light (e.g., Lotz et al.~2000;
Maraston \& Thomas 2000), suggesting that what we see directly
in NGC 5128 may be a general pattern for ellipticals.

In rough terms, what does this mean for the evolutionary history
of the galaxy?   A standard closed-box Simple Model of chemical evolution 
(which gives $dn/dZ \sim$ exp $(-Z/y)$ for an initial supply
of primordial $Z = 0$ gas and
where $y$ is the effective nucleosynthetic yield)
does well at matching the
upper end of the MDF ($Z \gtsim Z_{\odot}/3$)
but does very poorly at the low end.  
Such a model strongly overpredicts
the relative numbers of low$-Z$ stars in NGC 5128.

Within the context of these types of broad-brush
models, the most plausible alternative explanations
are (a) adopting a different IMF at early times, more strongly
weighted to high-mass stars, or (b) allowing continued infall of
unenriched gas {\sl during} the first phases of star formation
(an ``accreting-box'' model; see Binney \& Merrifield 1998).

The variable-IMF option cannot be definitively ruled out, but is an
arbitrary measure for which there is little or no compelling evidence
from the wide variety of present-day star-forming regions we see
around us now.
By contrast, option (b) -- ongoing gas accretion during the rapid
early phases of star formation -- is something that we would expect
to happen anyway for giant ellipticals at the centers of groups of
galaxies, regardless of other factors.

Once the infall option is introduced, we need to specify the
rate at which it happens.  Harris \& Harris (2001) 
define a simple ``delayed exponential'' infall model in which the
gas accretion rate $dM/dt$ is assumed to be constant for some initial time
interval $\tau_1$ and then declines exponentially over an $e-$folding
time $\tau_2$.  Both the initial infall rate and the fiducial times
$\tau_1, \tau_2$ are free parameters to be determined by the fit to the data.

Figure 2 displays a sample fit of one of these ``accreting-box'' 
models to the observed MDF for the outer-halo fields.  
The particular example shown has
$\tau_1=0$, i.e.~the gas accretion rate
declines exponentially right from the start.
The $e-$folding time $\tau_2$ here is set equal to 30 timesteps
where in our numerical model, 5\% of the ambient gas
is converted to stars in each timestep $\delta t$.  
A plausible and conservative assumption is that 
most of the accretion takes place over the first $\sim 3$ Gyr,
after which the gas composition
has built up to near-Solar levels.
Under these circumstances our average enrichment timestep
$\delta t$ would be about 30 Myr and the initial accretion rates
(and gas conversion rates) in the proto-elliptical are near
100 $M_{\odot}$ per year.

Almost any giant elliptical should be expected to grow, not just
by the direct conversion of gas to stars within its own
potential well, but by the later (and ongoing) accretion of
small satellites (e.g., C\^ot\'e et al. 2000).  How important is this
accretion in actuality?  Most stars in dwarf E galaxies, for example,
have metallicities in the range [m/H] = log $(Z/Z_{\odot}) < -0.7$.
The extreme lack of such low$-Z$ stars in NGC 5128 therefore 
places rather stringent constraints on the numbers of dwarf ellipticals
that could have been accreted.  It appears necessary to
conclude that the bulk of NGC 5128 was formed {\sl in situ} in the
classical way, or by mergers of much larger protogalactic clumps
which could hold on to most of their gas and thus build upward
to higher metallicity through continuing star formation and
enrichment.  

\section{Globular Clusters versus Field Stars}

What of the globular clusters?  Recent studies
(G.~Harris et al. 1992; Rejkuba 2001; Peng 2001)
show repeatedly that the MDF for the NGC 5128 clusters
(plotted now in its {\sl logarithmic} version as number per
unit [m/H]) has the bimodal form now well established
as the norm for giant ellipticals:  roughly equal numbers
of clusters in metal-poor and metal-rich modes, centered
approximately at log $(Z/Z_{\odot}) \sim -1.4$ and $-0.4$.
These are shown in Figure 3.

The high-metallicity peak for the clusters
matches up reasonably well with the peak of the stellar MDF
at log $(Z/Z_{\odot}) \simeq -0.4$,
but the two distributions are strikingly
different in their proportions of low-metallicity objects.  
It is just as obvious that the field stars,
{\sl not the clusters}, represent the MDF for the whole galaxy and
are the primary indicator of its overall star forming history.
Unless we can decipher the reason why the two distributions are
different, it casts considerable doubt on our perhaps-optimistic
claims from past years that the cluster MDF is a direct indicator
of the galaxy formation process.

Having the two full distribution
functions in hand for the first time within a giant elliptical allows
us to construct a new kind of graph:  the specific frequency (number
of clusters per unit halo light) {\sl as a function of metallicity}.
This function is shown in Figure 4.  
Whereas $S_N \sim 2-3$ for the metal-richer part of the MDF (quite
similar to the global average over all metallicities, since the vast
majority of the light comes from the high$-Z$ stars), it is in
the range $S_N \sim 10 - 20$ for the metal-poor component ([m/H] $< -1$).
In fact, $S_N$ at the low$-Z$ end is even higher than shown here,
because photometric scatter causes us to somewhat overestimate the
numbers of stars in that part of the CMD (Fig.~1).  There is thus
an {\sl order of magnitude} difference between the metal-poor and
metal-rich components in the ``efficiency'' with which they generated,
or acquired, globular clusters.

Values of $S_N \ga 20$ are almost unprecedented.  The only previously
known sites with such high ratios are a few of the smallest
nucleated-dwarf ellipticals (McLaughlin 1999; W.~Harris et al. 1998;
W.~Harris 2001).  A plausible interpretation of such objects (Durrell
et al. 1996; W.~Harris et al. 1998; McLaughlin 1999) is 
that they began their initial starburst with the globular clusters
forming earliest in the burst from the highest-density clumps.
Then, the stellar winds and supernovae from these first few objects
drove out most of the gas 
from its tiny potential well before it could be converted
to stars.  The net result was to leave behind a dE galaxy 
with only low-metallicity stars and an abnormally large number of clusters.

{\sl If} something similar happened at early times in the NGC 5128
halo, then there might be two possible answers to the $S_N(Z)$
trend shown in Fig.~5:

\noindent -- The halo stars with [m/H] $\la -1$ were essentially
all acquired by accretion of dE's, along with their (relatively
numerous) metal-poor globular clusters.  However, two worries 
are that (a) the observed level $S_N \sim 20$ is an
{\sl upper limit} for known dE's, whereas it is an {\sl average} for
the metal-poor halo in NGC 5128.  To make this work, we 
would have to require {\sl all} the accreted
satellites to be the very smallest types, which seems implausible,
or else to require that proto-dE's starting out within
much larger potential wells behaved differently from isolated ones.

\noindent -- Alternately, we might suppose that
the first round of star formation in NGC 5128 took place
in many widely distributed dwarf-sized clumps, each of which was located
within the larger proto-gE potential well.  These clumps were
already busy forming stars as they were simultaneously falling 
together to assemble the eventual giant galaxy.  Each clump could
have acted locally like a dE as described above, with a short intense
burst of star and cluster formation and ejection of most of the
gas into the larger potential well of the proto-gE.  This gas
later re-collected, initiating a second and more major round of
star formation that built the bulk of today's visible gE.
The postulated gas clumps are the supergiant molecular clouds or
SGMCs of Harris \& Pudritz (1994), or the dwarf-sized hosts of
Burgarella et al. (2001).
However, for this scenario to work,  
we must {\sl require} globular cluster formation to occur
early in any starburst. 

\section{Mass Distribution Functions:  Fact and Legend}

Much recent interest has centered on the mass distributions
of the old globular clusters and those of young clusters such as
the objects now known to be formed in starbursts
in large numbers.  
The globular cluster mass distribution (or equivalently
the luminosity distribution or LDF) is by tradition plotted as
number of objects per unit {\sl magnitude} (or log mass).  In this 
plane it is roughly Gaussian in form, and the growing body
of LDF observations continues to show how remarkably similar
it is from one galaxy to another.  By unfortunate contrast, the
LDF for young clusters is often plotted in the more physically
oriented plane of number per unit {\sl luminosity} (or mass),
where it appears more or less like a power law.  The two planes
are directly related ($dN/dm_V \sim dN/d{\rm log}M \sim M\cdot dN/dM$
for magnitude $m_V$ and mass $M$), but an astonishing amount
of discussion in the recent literature has emphasized the 
``differences'' between old and young clusters which in fact
boil down to this trivial issue of notation.  

In the linear plane ($dN/dM$ or its observable surrogate $dN/dL$),
a typical LDF for
old-halo clusters looks like the one shown in Figure 5.  The particular
case shown is for the
prototypical giant elliptical M87.  Above the $\sim 10^5 M_{\odot}$
point which marks the traditional GCLF ``turnover'', a single
power-law slope $dN/dL \sim L^{-1.8 \pm 0.2}$ is consistently found
to match the globular cluster systems in most or all galaxies.
For young-cluster systems (the best-studied example being the
Antennae clusters; Whitmore et al. 1999) the same power law
slope is seen.  Dynamical evolution over the first $1-2$ Gigayears has 
very likely been responsible for carving away the low-mass end 
($< 10^5 M_{\odot}$), but the higher-mass end maintains something close
to its original shape (cf. Vesperini 2001; Fall 2001).

But there is one important contrast between the Antennae and M87
clusters which has not drawn enough attention.  Their median
masses are quite different.  The Antennae merger is forming
star clusters at a characteristic mass scale {\sl about
one order of magnitude smaller} than the traditional globular
clusters in M87 and elsewhere.  The Antennae merger certainly contains
a few of the ultra-large GMCs of $\gtsim 10^8 M_{\odot}$ 
that we are looking for 
(Wilson et al. 2000), but the vast majority of its star-forming
sites are within much smaller clouds which will not leave
behind massive star clusters.  The suggestion hidden
in this comparison is that giant ellipticals like M87
formed by the assembly of GMCs that were, on average, much 
larger than we now see in the Antennae.
It is all the more remarkable
that the power-law shape of the mass distribution seems to be,
roughly at least, independent of mass scale.

Various plausible alternatives have been suggested for building
the basic form of the initial mass distribution function, ranging from
collisional growth of cloudlets within GMCs (McLaughlin \& Pudritz 1996)
to turbulence spectra (Elmegreen \& Efremov 1997).  It is worth 
emphasizing that the collisional-growth model can provide 
{\sl quantitative} matches to the observed mass distributions in
a natural way as long as the cloud lifetime against internal
star formation is about an order of magnitude longer than the
cloud/cloud collision time (see Harris \& Pudritz 1994; McLaughlin
\& Pudritz 1996 for thorough discussion).  Fortunately, this
requirement is likely to be
automatically true if turbulence and magnetic field are the
dominant sources of internal pressure.

I will close this review with a few remarks on the role of
the GMCs within which stars and star clusters form.  Harris \& Pudritz
(1994) postulated that reservoirs of gas $10^8 -10^9 M_{\odot}$ big
-- ``supergiant'' molecular clouds or SGMCs -- would be needed in order
to build globular-cluster-sized protoclusters of $10^5 - 10^6 M_{\odot}$,
arguing from the empirical evidence that the biggest star clusters
formed in a given GMC attain only a few thousandths of the whole
GMC mass.  Clearly, the very high star formation efficiency $e \sim 0.5$
necessary to form a gravitationally bound cluster is a rather rare
occurrence.  The emergent mass spectrum of their embedded star clusters
would, in this scenario, be a miniature replica of the host GMC
mass distribution if the characteristic ratio of protocluster
mass to parent cloud mass $\eta \equiv M_{cl}/M_{GMC}$ is independent
of mass scale.  Lastly, the total {\sl number} of clusters, and hence
the specific frequency, is set by the number of clusters
a given GMC can produce; again, strictly empirical evidence suggests
that half a dozen bound clusters per GMC is typical.

Current work is filling in more of the comprehensive picture.
Weil \& Pudritz (2001) use N-body modelling within a cosmological
CDM simulation to explicitly
follow the buildup of gas clouds within the dark-matter potential
well of a protogalaxy.  They find 
that hierarchical growth creates a GMC cloud
mass spectrum of the form $dN/dM \sim M^{-1.7\pm 0.1}$.  Furthermore,
the ``top end'' of the mass distribution reaches $10^9 M_{\odot}$
quite quickly (by redshift $z \simeq 3.5$).  These features strongly
resemble the dwarf-sized SGMCs that we have argued must be the
necessary host sites for globular cluster formation in the early
galaxy.  They have also been cross-identified with the damped Ly$\alpha$
systems by Burgarella et al. (2001; see also Kissler-Patig 2001).

These simulations are both intriguing and encouraging, but we are
still far from tying together the complete story.  For example,
the Weil/Pudritz models trace only what happens to the gas
(the prescriptions for star formation are deliberately switched off);
and we do not understand the 
crucial mass ratio parameter $\eta \sim 10^{-3}$ 
which sets the mass scale of the clusters within a given GMC.
Nevertheless, there are grounds for genuine optimism that we
are beginning to piece together a more comprehensive history
for these fascinating objects.

\clearpage

\figcaption{Color-magnitude array for the NGC 5128 halo stars in the
21-kpc and 31-kpc target fields, from Harris \& Harris (2000).  
The red-giant model tracks extending up to the helium
flash extend from heavy-element abundances of $0.005 Z_{\odot}$ up to
almost $3 Z_{\odot}$.}

\figcaption[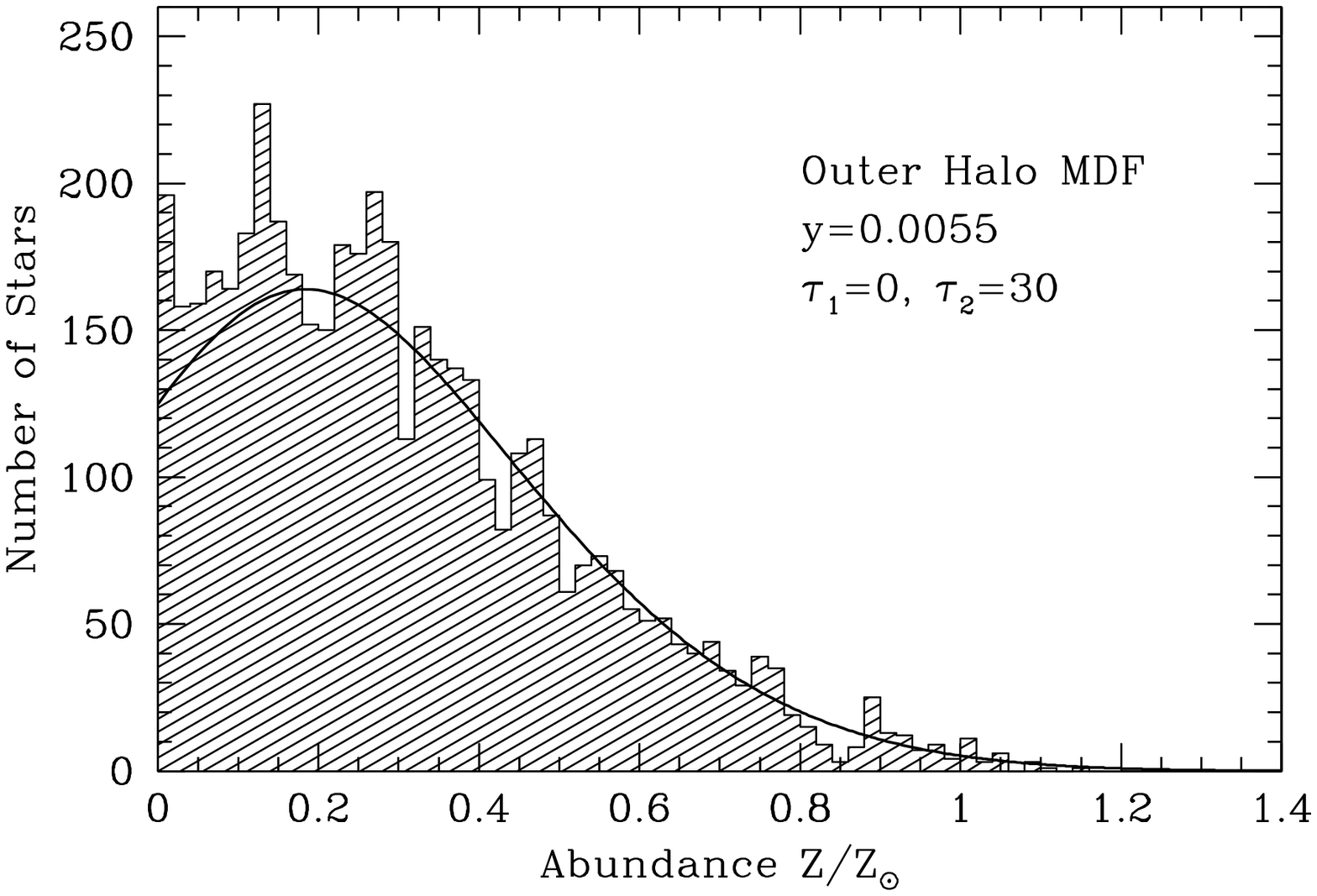]{Metallicity distribution function (MDF) for the outer
halo stars in NGC 5128, from Harris \& Harris (2000).
Note that the MDF is plotted in its linear form as number of
stars per unit heavy-element abundance $dn/dZ$.
The solid line shows a model of chemical evolution starting 
with $Z=0$ gas and allowing gas infall at an exponentially
decaying rate (see text).
}

\figcaption[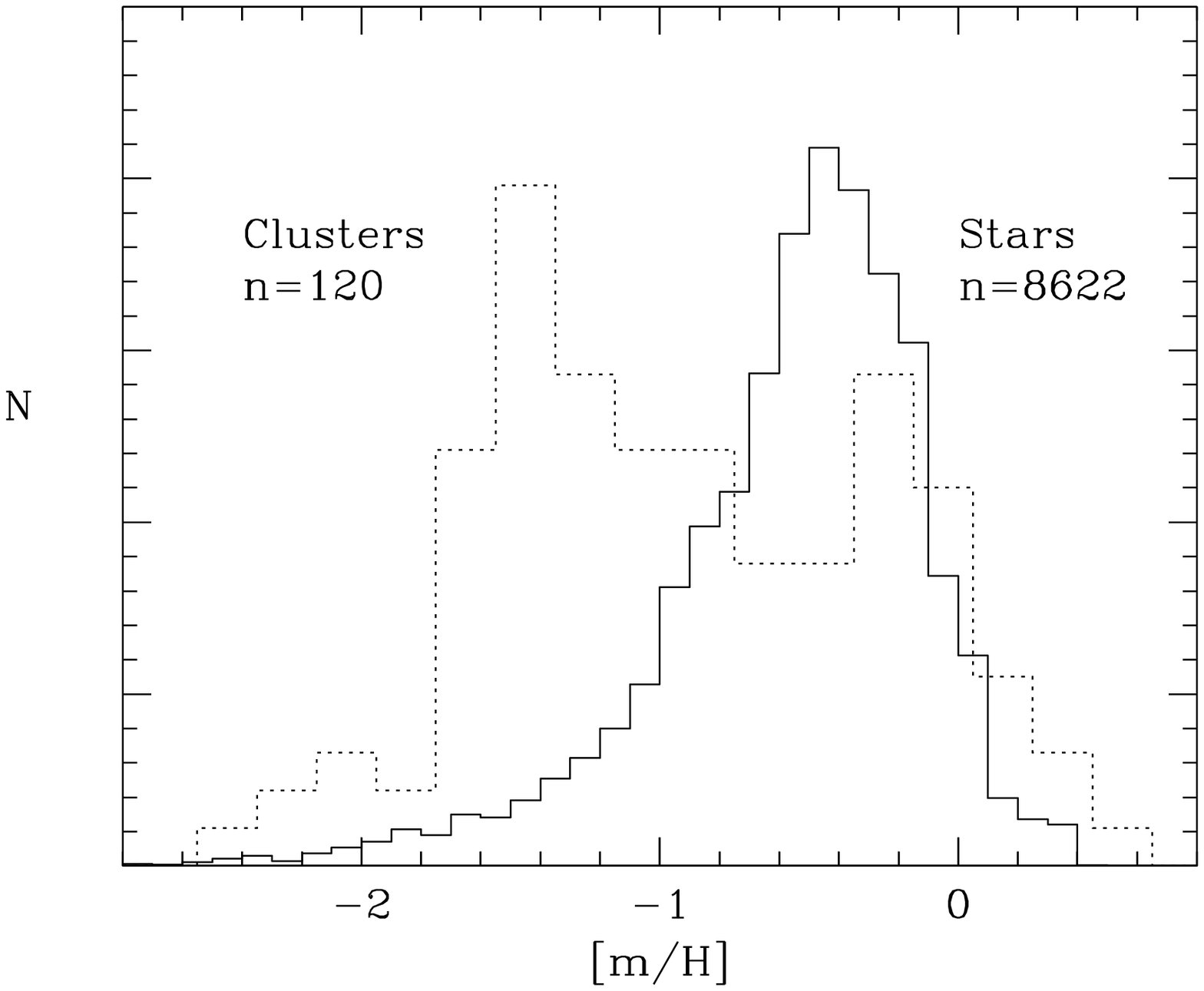]{Metallicity distribution functions for the globular
clusters and halo field stars in NGC 5128.  The clusters (dotted
line; from the combined data of G.~Harris et al. 1992 and Rejkuba
2001) have the characteristic bimodal form seen in most giant
ellipticals. The stars belong almost entirely to the metal-rich ``mode''.
}

\figcaption[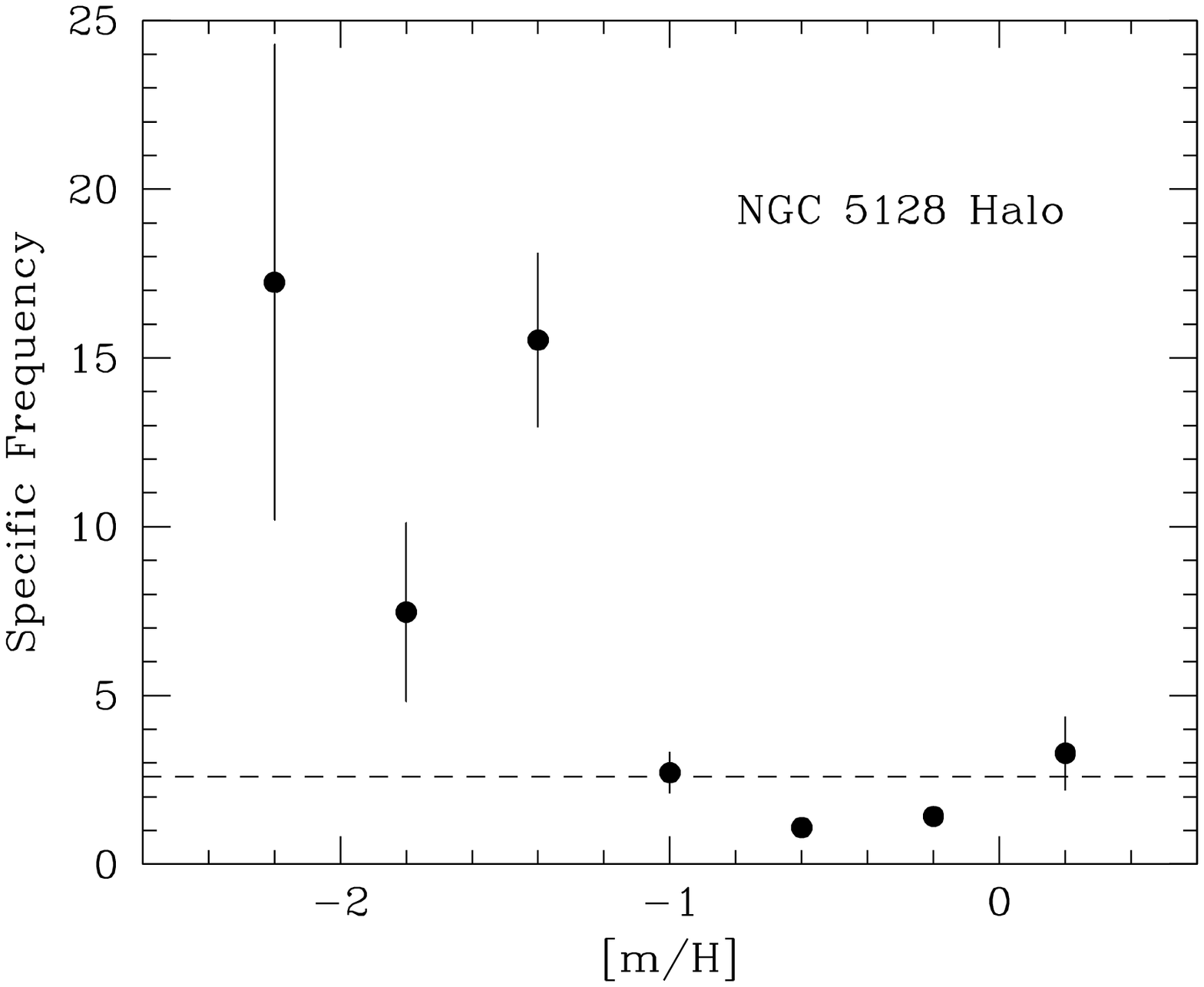]{Specific frequency versus metallicity for the halo
of NGC 5128 (from Harris \& Harris 2001).  The
total $S_N$ over the whole galaxy (dashed line) is at $S_N = 2.6$.  
}

\figcaption[harris2_fig4.ps]{Luminosity distribution function (LDF) for globular
clusters in M87 (data adapted from Kundu et al.~1999).
Note here that the LDF is plotted in its {\sl linear} form
as number of clusters per unit $10^4 L_{\odot}$.  
The {\sl solid line} is a theoretical mass spectrum 
(McLaughlin \& Pudritz 1996).  The specific
model shown is for a timescale ratio $\beta = 23$ (star formation
time divided by cloud collision time) and for a cloud lifetime
varying as $\tau \sim M^{-1/2}$.
}

\end{document}